\documentclass[12pt]{article}
\usepackage[utf8]{inputenc} 
\usepackage[T1]{fontenc}    
\usepackage{nicefrac}       
\usepackage{microtype}      
\usepackage{graphicx}
\usepackage{amsmath, amsthm, amsfonts}

\usepackage{algorithmic, algorithm}
\usepackage[font=small]{caption}
\usepackage{wrapfig}
\usepackage[scaled]{helvet}
 
\DeclareMathOperator{\diff}{d}
\usepackage{authblk}
\title{New Light on Cortical Neuropeptides and Synaptic Network Plasticity}
\author[1]{Stephen J Smith\thanks{Corresponding Author: stephens@alleninstitute.org}}
\author[1]{Michael Hawrylycz\thanks{mikeh@alleninstitute.org}}
\author[2]{Jean Rossier\thanks{jean@rossier.paris}}
\author[1]{Uygar S\"umb\"ul\thanks{uygars@alleninstitute.org}}
\affil[1]{Allen Institute for Brain Science, 615 Westlake Ave N, Seattle WA, USA}
\affil[2]{Neuroscience Paris Seine, Sorbonne Université, Paris, France}
\date{}
\begin{document}
\maketitle

\section*{Highlights}
\begin{itemize}
\item New findings expand the evidence for involvement of local neuropeptide signaling in cortical synaptic network plasticity.
\item Single-cell transcriptomes indicate that every cortical neuron secretes at least one neuropeptide and displays at least one neuropeptide receptor.
\item Transcriptomic neurotaxonomies predict cortical peptidergic network architectures.
\item New genetic tools open transcriptomic network predictions to experimental test.
\item New genetic tools open neuropeptide roles in synaptic network plasticity to experimental and theoretical exploration
\end{itemize}

\begin{abstract}
    Neuropeptides, members of a large and evolutionarily ancient family of proteinaceous cell-cell signaling molecules, are widely recognized as extremely potent regulators of brain function and behavior. At the cellular level, neuropeptides are known to act mainly via modulation of ion channel and synapse function, but functional impacts emerging at the level of complex cortical synaptic networks have resisted mechanistic analysis. New findings from single-cell RNA-seq transcriptomics now illuminate intricate patterns of cortical neuropeptide signaling gene expression and new tools now offer powerful molecular access to cortical neuropeptide signaling. Here we highlight some of these new findings and tools, focusing especially on prospects for experimental and theoretical exploration of peptidergic and synaptic networks interactions underlying cortical function and plasticity.
\end{abstract}

\begin{center}
\textcopyright 2020, This manuscript version is made available under the CC-BY-NC-SA 4.0 license. \\
http://creativecommons.org/licenses/by-nc-sa/4.0/
\end{center}

\newpage
\section*{Introduction}
Neuropeptides are diffusible cell-cell signaling molecules that act in brain as powerful regulators of social valence, sleep, appetite, anxiety, stress response, pain perception and memory~\cite{meriney2019synaptic, gotzsche2016role, deutch2014nonclassic, borbely2013neuropeptides, van2012neuropeptide, burbach2011neuropeptides}. Many neuropeptides were first discovered as neuroendocrine hormones regulating growth, reproduction, gut function or other bodily processes, but later shown to also signal strictly between clearly non-neuroendocrine neurons~\cite{van2012neuropeptide}. Others, such as the opioid neuropeptides, were discovered during molecular and physiological explorations focused on central nervous system perceptual and cognitive impacts~\cite{simon1973search}. 

Well over one hundred genes encoding primary neuropeptide signaling proteins, both precursors and receptors, are present in the human genome. A high degree of sequence conservation across these genes - spanning the entire animal kingdom - argues strongly for very ancient evolutionary origins of peptidergic signaling~\cite{elphick2018evolution, jekely2013global, varoqueaux2017getting}. There is even evidence that neuropeptides may have coordinated the behavior of slow-moving, ancestral metazoans before the advent of neurons and specialized, fast synaptic connections~\cite{varoqueaux2017getting, smith2019coherent}. Though most animal species today rely primarily upon fast synaptic transmission using recycling small-molecule neurotransmitters (with rapid re-uptake or degradation) for their speedy behaviors, neuropeptides remain critical to adaptive nervous system function, acting via slower biochemical coupling as modulators of ion channel and synaptic function~\cite{van2012neuropeptide, liguz2016somatostatin, nadim2014neuromodulation, mccormick2014editorial, bargmann2012beyond}. It now appears likely that most (and perhaps all) present-day neurons both secrete and respond to at least one neuropeptide, in addition to a small-molecule synaptic neurotransmitter, and that such co-release may be critical to synaptic network function~\cite{van2012neuropeptide, cropper2018peptide, granger2017multi, nusbaum2017functional, svensson2019general}.

Transcriptomic and physiological evidence now points to a view of mammalian cerebral cortex as a stack of fast synaptic and slow peptidergic modulatory networks that interconnect neurons of highly differentiated synaptic and neuromodulatory connection affinities.  Here we’ll summarize some of this evidence and remark upon a few of the new molecular, anatomical, physiological and theoretical tools now bringing new light to these highly diverse cortical networks and their interactions. Finally, we’ll mention ways that interests converging from neuroscience and computer science disciplines may deepen our understanding of memory engram formation by deep synaptic networks, be they natural or artificial. We begin here with a brief overview of the present understanding of peptidergic signaling mechanisms.

\section*{The neuropeptide signaling canon in brief}
Active neuropeptide (NP) molecules (small peptides usually consisting of 3-36 amino acids) are produced within source neurons by enzymatic fragmentation and covalent modification of neuropeptide precursor protein (NPP) gene products, stored in large, dense-cored vesicles and then secreted in response to calcium influx during neuronal activity~\cite{meriney2019synaptic, deutch2014nonclassic, burbach2011neuropeptides}. NP molecules, once secreted, persist for many minutes in brain interstitial spaces, long enough for paracrine diffusion over distances embracing hundreds of potential target neurons~\cite{meriney2019synaptic, alpar2019hypothalamic, nassel2009neuropeptide, chini2017action, leng2008neurotransmitters}. In contrast to synaptic signaling, where small molecule transmitter actions are restricted by rapid re-uptake or degradation processes to just one very closely apposed postsynaptic neuronal site, a neuropeptide is “broadcast” over much longer, though still limited, ranges set by release site anatomy, quantities secreted, diffusion physics and slow degradation processes. Such paracrine neuropeptide signaling may nonetheless precisely address specific neurons within a broadcast volume: only neurons expressing receptors “cognate to” (i.e., selective for) a particular secreted peptide molecule will respond~\cite{meriney2019synaptic, deutch2014nonclassic, van2012neuropeptide}. Consider an analogy with radio or television broadcasting, where many receivers generally exist within a given broadcast area, but only those sets tuned to the specific frequency of a particular station respond to the broadcast. Neuropeptide signal connectivity is thus surely determined in large part by cell-specific patterns of differential ligand and receptor gene expression. 

Most neuropeptide receptors are encoded by members of the very large superfamily of G-protein-coupled receptor (GPCR) genes~\cite{meriney2019synaptic, deutch2014nonclassic}. GPCRs are selective, high-affinity receptors distinguished by characteristic seven-transmembrane-segment atomic structures and signal transduction involving heterotrimeric G-proteins (hence the terminology)~\cite{weis2018molecular, syrovatkina2016regulation}. Most GPCR genes encode olfactory receptors~\cite{zhang2002olfactory, buck1991novel}, but mouse and human genomes still comprise over 350 genes encoding GPCRs selective for diverse endogenous ligands. At least one hundred of these are neuropeptide-selective (NP-GPCRs), while dozens more are selective for other widely studied neuromodulators including dopamine, serotonin, norepinephrine and endocannabinoids~\cite{vassilatis2003g}. 

Though GPCRs are highly diverse in ligand specificity, they are much less diverse in downstream signaling actions. While GPCR signaling has many additional facets and complexities~\cite{meriney2019synaptic, gurevich2019gpcr, hille2015phosphoinositides, borroto2019oligomeric, wacker2017ligands}, the primary actions of most known neuronal GPCRs are mediated by the intracellular second messengers cyclic AMP, ionic calcium, and/or phosphoinositide (PI) lipid metabolites. These second messengers, in turn, act via regulation of protein kinases, directly upon ion channels, or upon gene expression to impact neuronal firing and synaptic transmission on a wide range of time scales. Primary effects of the GPCRs expressed in cortex fall into just three major categories distinguished by G-protein alpha subunit (G$\alpha$) family: the Gi/o family (i/o) inhibits cAMP production, the Gs family (s) stimulates cAMP production, and the Gq/11 family (q) stimulates phospholipase C activity to generate PI metabolites and amplify calcium signaling~\cite{smith2019single}. For most NP-GPCR genes, the primary G$\alpha$ family (e.g., i/o, s or q/11) is now known~\cite{armstrong2020iuphar} and provides a useful first-order prediction of the encoded GPCR’s signal transduction impact. Though this basic trichotomy is certainly an oversimplification, such predictions are useful to form presently fragmentary information about actions of neuropeptides and diverse other GPCR-mediated modulatory ligands into experimentally testable hypotheses. Broad conservation of GPCR sequences also makes it reasonable to tentatively generalize mechanisms and impacts of neuropeptide signaling across different neuropeptides, brain regions and animal phyla~\cite{elphick2018evolution,jekely2018long}. We’ll now touch upon new opportunities to evaluate and leverage such generalizations.
\begin{figure}
    \centering
    \includegraphics[width=0.99\textwidth]{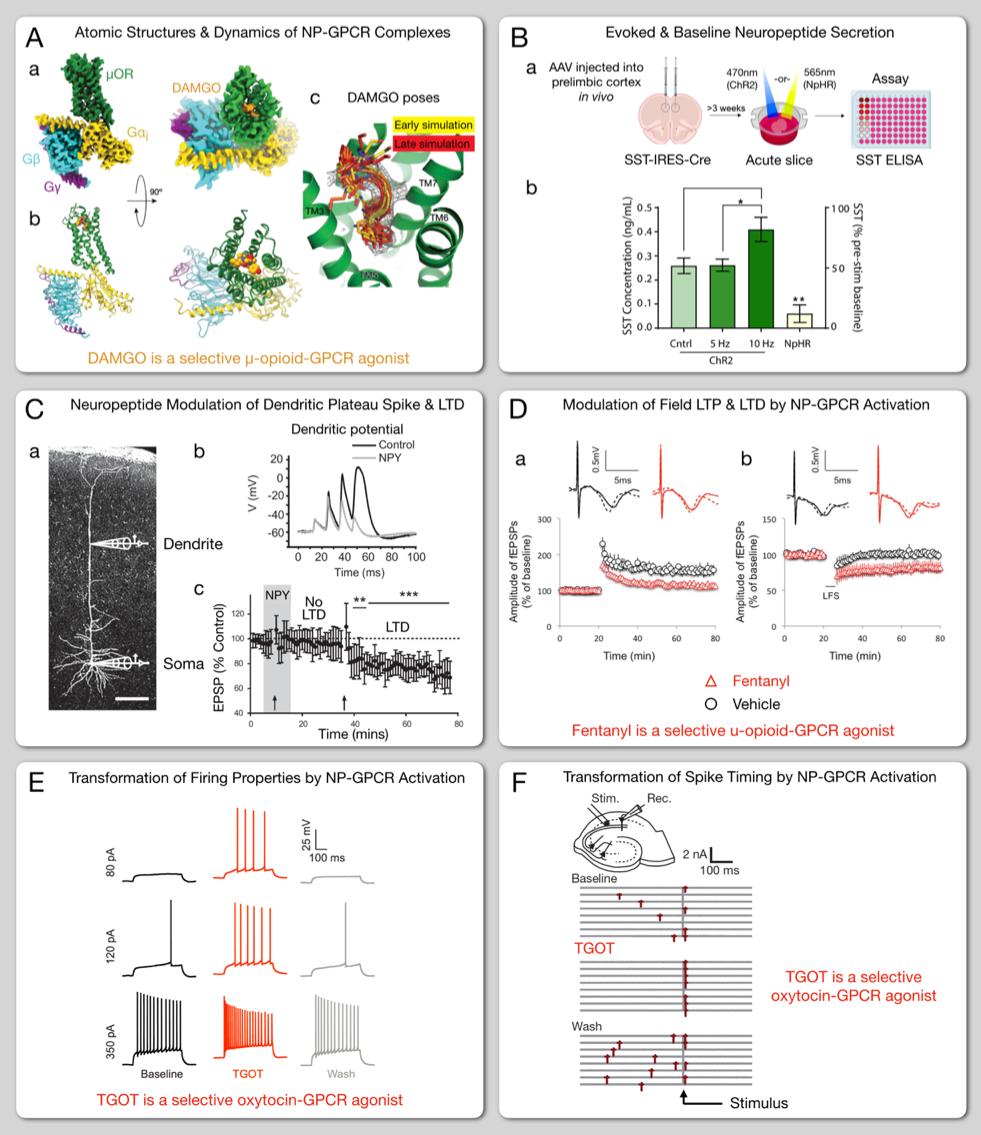}
    \caption{\small {\bf New results on neuropeptide signaling and synaptic network plasticity.} {\bf A.} Atomic structure and molecular dynamics a mu-opioid-GPCR/Gi heterotrimer complex bound to the selective agonist DAMGO~\cite{koehl2018structure}. (a) Orthogonal views of the cryo-EM density map of this complex colored by subunit, as indicated. (b) Model of the same complex in the same views and colors as in a. (c) A frame from every 100 ns of a 1 $\mu$s molecular dynamics simulation shows that the first four residues of DAMGO (bottom) are stable, whereas the C-terminal (top) is more dynamic.
{\bf B.} Measurement of somatostatin (SST) release from a specific subset of mouse cortical interneurons~\cite{dao2019vitro}. (a) Experimental design using type-specific}
\label{fig1}
\end{figure}
\setcounter{figure}{0}
\begin{figure}
    \renewcommand{\thefigure}{\arabic{figure} (ctd)}
    \caption{\small
     driver lines~\cite{daigle2018suite} and anatomic localization via viral vectors~\cite{haery2019adeno} to target optogenetic tools precisely. (b) Results showing that stimulation of high-frequency firing accelerates release of SST while suggesting that SST is also released by spontaneous baseline activity. 
{\bf C.} Dendritic patch recordings (a) show that NPY suppresses both (b) plateau spiking and (c) triggering of LTD (a form of STDP~\cite{turrigiano2017dialectic}) in rat somatosensory cortex~\cite{hamilton2013modulation}, suggesting that modulation of dendritic plateau spiking by NPY may underlie NPY modulation of STDP. {\bf D.} Activation of NP-GPCRs by fentanyl (a selective agonist for the mu-opioid NP-GPCR) modulates STDP in rat hippocampal cortex area CA1~\cite{tian2015effect}, suppressing triggering of LTP (a) and enhancing triggering of LTD (b).
{\bf E.} Electrophysiological evidence for a profound cellular effect of the neuropeptide oxytocin in mouse hippocampal cortex area CA2~\cite{tirko2018oxytocin}. The oxytocin analogue TGOT transforms pyramidal cells firing properties. Such neuropeptide transformation of spike firing is likely to translate into modulation of STDP.
{\bf F.} Electrophysiological evidence for a profound network effect of the neuropeptide oxytocin in mouse hippocampal cortex area CA1~\cite{owen2013oxytocin}. TGOT reversibly enhances stimulus-related spike synchrony by suppressing spontaneous firing and enhancing elicited firing in pyramidal cells. Such neuropeptide transformation of synchrony is also likely to translate into modulation of STDP.}
\end{figure}

\section*{New light on mechanisms and effects of cortical neuropeptide signaling}
Our understanding of cortical synaptic network plasticity is now benefitting from advances in elucidating basic molecular mechanisms and effects of neuropeptide signaling. One overarching advance, exemplified in Fig. 1A, has been the determination of atomic structures of a rapidly growing number of GPCRs~\cite{weis2018molecular, safdari2018illuminating, koehl2018structure}, coupled with structure-based molecular dynamic simulations of interaction of GPCRs with ligands and heterotrimeric G proteins~\cite{durdagi2019current}. Besides providing major insights into the structural basis of NP-GPCR function, these new results are now enabling the development of genetically encoded sensors and effectors targeting specific neuropeptide ligand-receptor systems, as discussed in the section “New physiological tools“ below.

In combination with new genetically encoded sensors and effectors, neuron-type-specific gene expression driver lines and viral vector injection methods, as exemplified in Fig. 1B, and emerging gene editing methods now provide means to target neuropeptide signaling in specific genetically and anatomically defined neuron subsets within intact cortical tissues. Such means seem certain to help resolve troublesome uncertainties as to function that have resulted until now from the well-established but poorly defined presence of dauntingly large numbers of peptidergic ligands and receptors in nearly every brain region. Figures~\ref{fig1}C-F represent ways that more traditional electrophysiological and pharmacological tools also continue to provide new insight into neuropeptide signaling impacts, focusing especially on indications of neuropeptide involvement in modulation of synaptic plasticity. Combinations of new genetic tools with advanced physiological and pharmacological methods are well matched to the substantial remaining challenges of understanding peptidergic modulation of cellular, synaptic and network function and plasticity.

\section*{New light on the patterning of neuropeptide precursor and receptor gene expression}
Neuron type diversity, first recognized by pioneering nineteenth century microscopists, is a fundamental element of nervous system organization~\cite{bates2019neuronal, fishell2019interneuron, gatto2019neuronal, tosches2019evolution, zeng2017neuronal}. Over recent decades, differential immunostaining for various neuropeptides (e.g., vasoactive intestinal peptide, somatostatin, neuropeptide Y, substance P, and cholecystokinin) has come into wide use for discriminating anatomically and functionally distinct neuron types~\cite{l2017neurochemical}. More recently, in situ hybridization and microarray gene expression data have established that mRNA transcripts encoding these five NPPs and many other NPP and NP-GPCR genes are expressed in large numbers of cortical neurons~\cite{lein2007genome, sugino2006molecular}. Expression data combining whole-genome depth with single-cell resolution have been lacking, however, and almost unimaginable until recently. This lack has hindered exploration of the full extent of cortical neuropeptide gene expression and made it difficult to design robust experiments to develop and test cogent hypotheses regarding the network consequences of cortical neuropeptide signaling.
Single-cell RNA-seq methods have now revolutionized the study of differential gene expression in brain~\cite{fishell2019interneuron, zeng2017neuronal, tasic2018shared, paul2017transcriptional, huang2019diversity}. We recently reported a neuropeptide-focused analysis~\cite{smith2019single} of a body of such data acquired from over 23,000 individual mouse cortical neurons~\cite{tasic2018shared}. Our analysis supports the surprising conclusion that all (or very nearly all) cortical neurons express at least one NPP gene at an extremely high level, and at least one NP-GPCR gene. The extreme abundance of NPP transcripts in most cortical neurons can be considered prima facie evidence that the corresponding NP products are produced, secreted and functionally important~\cite{smith2019single}. Furthermore, the vast majority of individual cortical neurons actually express more than one (sometimes as many as ten!) distinct NPP and NP-GPCR genes.

\begin {table}
\begin{center}
\includegraphics[width=0.99\textwidth]{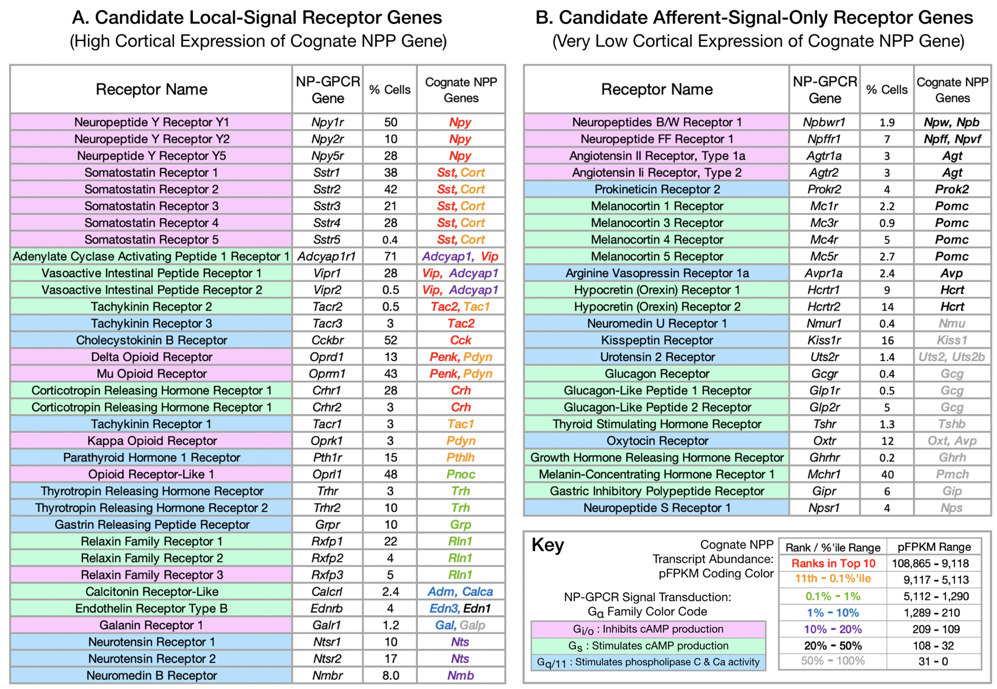}
\end{center}
  \caption{\small {\bf 58 NP-GPCR genes differentially expressed in mouse cortical neurons.} 
Two columns differentiate 58 NP-GPCR cortically expressed genes into two classes based on expression patterns of their cognate NPP genes, as measured by single-cell RNA-seq (Adapted from~\cite{smith2019single}). 
{\bf A.} 34 cortically expressed NP-GPCR genes for which high cortical expression of cognate NPP genes suggests a strong possibility of local signaling between pairs of cortical neurons.
{\bf B.} 24 cortically expressed NP-GPCR genes for which no known cognate NPP gene is expressed appreciably in cortex, suggesting that functional signaling likely reflects release from afferent axons projecting into cortex from extracortical (e.g. hypothalamic) sites of cognate NPP gene expression.
Tinted fills in the “receptor name” columns indicate G$\alpha$ signal transduction family membership for each NP-GPCR (see Key). Fractions of neurons expressing these NP-GPCR genes were calculated in FPKM units from the data of~\cite{tasic2018shared}, analyzed and tabulated as described in~\cite{smith2019single}. Single-cell peak FPKM (pFPKM) values for cognate NPP genes were evaluated similarly and are represented here by the brackets indicated as typeface colors (see Key). The cognate NPP cutoff pFPKM separating NP-GPCR genes in Column A (top quintile of pFPKM values, across all protein-coding genes) from those in Column B (below top quintile), though somewhat arbitrary, is based on the large gap dividing the two cognate NPP ranges (top quintile cognate NPP pFPKM values range from 108,865 down to 112, median $\sim$4,500, while the highest cognate NPP pFPKM value below cutoff is 76).}
\label{table1}
\end{table}
Table~\ref{table1}, adapted from reference~\cite{smith2019single}, tabulates a total of 58 NP-GPCR genes that are expressed at substantial levels in mouse sensory and motor cortical areas, along with identities of corresponding cognate NPP genes (i.e., NPP genes for which the expected neuropeptide product matches the ligand selectivity of a given NP-GPCR). Of the 58 NP-GPCR genes, the 34 in column A are cognate to an NPP gene expressed at very high levels in the same cortical areas, suggesting the possibility of locally intracortical neuropeptide signaling. This suggestion is especially compelling for the GABAergic interneurons, since most have only short local axons, making it unlikely that their secreted NP products could act at very remote sites. Conversely, very low expression within a given cortical locale of NPP genes cognate to the 24 column B NP-GPCR genes suggests that the corresponding 24 receptors will never activate in response to locally generated neuropeptide. The column B NP-GPCRs nonetheless remain strong candidates for activation by neuropeptides released by afferent axons projecting from distant subcortical or cortical structures, where the relevant cognate NPP genes are in fact expressed.
Single-cell RNA-seq data have also enabled the precise data-driven clustering of large and diverse neuron populations into neurotaxonomies representing tractably small numbers of neuron classes, subclasses and types~\cite{fishell2019interneuron, tosches2019evolution, zeng2017neuronal, tasic2018shared, paul2017transcriptional, huang2019diversity, mukamel2019perspectives}. Such transcriptomic neurotaxonomies become very useful if one assumes that the individual neurons within each type cluster share similar single-neuron physiological dynamics, patterns of synaptic and neuromodulatory connectivity to other neuron types, and play similar roles in adaptive network function. Though further critical examination of this neurotaxonomy premise is necessary and ongoing~\cite{fishell2019interneuron, mukamel2019perspectives}, it seems reasonable and very likely to reward further experimental and theoretical pursuit~\cite{zeng2017neuronal, paul2017transcriptional}.
\begin{figure}
    \centering
    \includegraphics[width=0.99\textwidth]{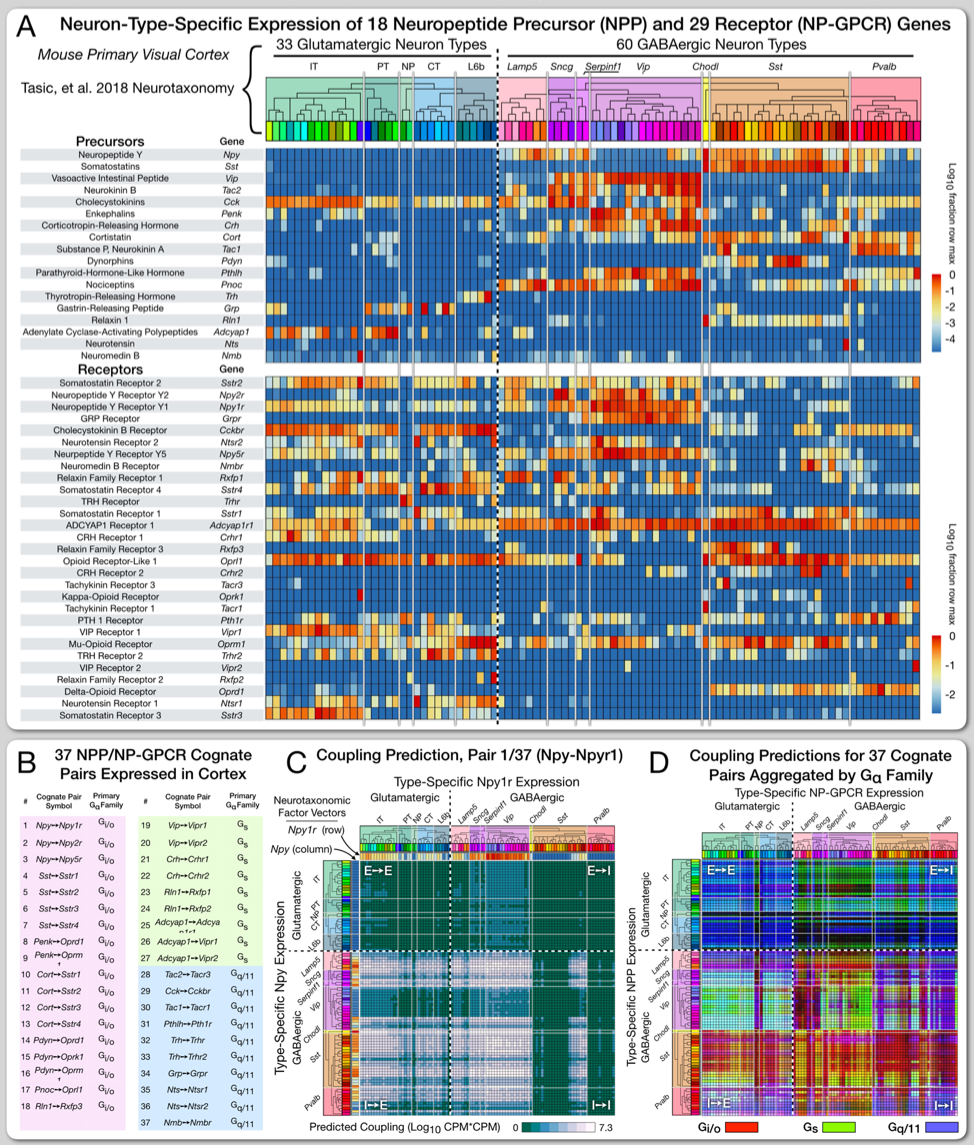}
\caption{\small {\bf Predictions of peptidergic network coupling from single-cell RNA-seq data.} (Figure adapted from~\cite{smith2019single}).
{\bf A.} “Heat map” matrices representing single-cell RNA-seq expression profiles of 18 neuropeptide precursor (NPP) and 29 neuropeptide-selective receptor (NP-GPCR) genes (rows) across all 33 glutamatergic and 60 GABAergic neuron types distinguished in mouse primary visual cortex by transcriptomic neurotaxonomy~\cite{tasic2018shared} (columns).
{\bf B.} The 37 cognate pairs embraced by the 18 NPP and 29 NP-GPCR genes of panel A. A column so designated and pastel background tints (same as Table~\ref{table1}) highlight primary G$\alpha$ transduction families for each pair.}
\label{fig2}
\end{figure}
\setcounter{figure}{1}
\begin{figure}
    \renewcommand{\thefigure}{\arabic{figure} (ctd)}
    \caption{\small {\bf C.} Matrix predicting neuron-type-specific peptidergic coupling on the basis of expression profiles for the Npy-Npyr1 cognate pair (\#1 of the 37 pairs listed in panel B. This matrix is the outer product of a column vector representing the taxonomic profile of Npy gene expression and a row vector representing Npyr1 expression. Matrix is divided into four quadrants according to major neuron class (E: glutamatergic; I: GABAergic). Analogous matrices for the other 36 pairs are presented in reference~\cite{smith2019single}.
{\bf D.} Coupling predictions aggregating all 37 cognate pairs based on $G\alpha$ family and thus expected second-messenger impacts, to facilitate experimental exploration using genetically encoded sensors of intracellular calcium and cyclic AMP like those sampled in Figure~\ref{fig3}.}
\end{figure}
Figure~\ref{fig2} represents results, described more fully in reference~\cite{smith2019single}, where we examined patterns of neuropeptide precursor and receptor gene expression in mouse neocortex through the powerful prism of transcriptomic neurotaxonomy. Fig. 2A illustrates our finding that expression profiles of 18 NPP and 29 NP-GPCR genes suggest that each transcriptomic neuron type expresses its own unique combination of these 47 NP genes. This finding is consistent with the well-established utility of neuropeptides as molecular markers of neuron type~\cite{l2017neurochemical} but extends to a considerably deeper neurotaxonomy. Indeed, we have presented a machine-learning analysis indicating that a neuron-type clustering based solely on differential expression of these 47 genes matches deep clustering results based on genome-wide clustering~\cite{tasic2018shared} and does so with extraordinary precision in comparison to other, similarly small gene sets~\cite{smith2019single}.

Recalling our radio broadcast metaphor, a neuron expressing a given NPP gene and another neuron expressing a cognate NP-GPCR gene can be cast as a directed communication channel, where “tuning” of both the “transmitter” and the “receiver” are established by each neuron’s gene expression profile. Broadcasts on different channels might communicate separate messages simultaneously to receivers tuned to separate channels within the same geographical reception area. The 18 NPP and 29 NP-GPCR genes whose differential expression profiles are represented in Fig. 2A comprise the 37 different cognate NPP/NP-GPCR pairs tabulated in Fig. 2B, where each cognate pair can be considered a distinct communication channel, with addressing fixed by gene expression profiles. Figure~\ref{fig2}C is a prediction of one 93 x 93 coupling matrix or “molecular connectome” derived from expression profiles for one single cognate NPP/NP-GPCR pair, Npy-Npyr1, amongst the 93 neuron types represented in Fig. 2A. Since the data of Fig. 2A actually cover 37 analogous cognate pairs, they actually predict of a stack of 37 neuron-type-based peptidergic coupling matrices (all 37 of which are laid out for both mouse visual and motor cortical areas in reference~\cite{smith2019single}). Fig. 2D aggregates the entire stack of 37 coupling matrices for visual cortex, factored by the three G$\alpha$ signal transduction families and their expected second-messenger impacts~\cite{smith2019single}. Coupling matrices like those of Fig. 2C and 2D can be understood as experimentally testable predictions of dense peptidergic modulatory networks. The experimental work needed to test such predictions and explore their ramifications is now feasible given the ongoing emergence of tools like those summarized in the following section.

\section*{New physiological tools}
\begin{figure}
    \centering
    \includegraphics[width=0.99\textwidth]{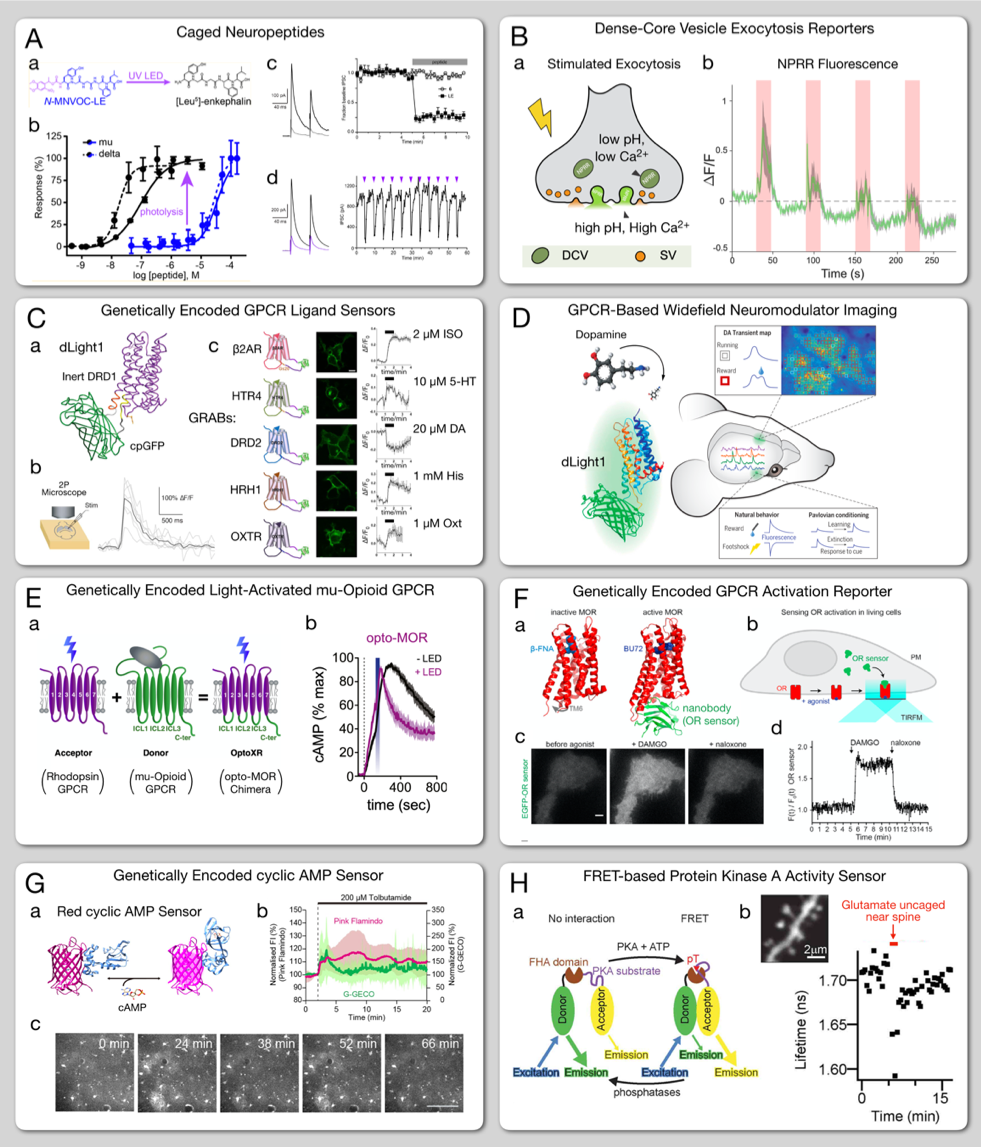}
\caption{\small {\bf New physiological tools to probe cortical neuropeptide signaling.} Used with neuron-type-specific targeting (e.g., via Cre lines or viral vectors), such probes are poised to enable tests of type-specific coupling matrix predictions (as exemplified in Figure~\ref{fig2}) and many other experimental explorations of neuropeptide roles in nervous system function. {\bf A.} (a) N-MNVOC-LE, a UV-LED-activatable caged form of leu-enkephalin, an endogenous neuropeptide agonist for both mu and delta opioid NP-GPCRs. (b) Photolysis increases N-MNVOC-LE activity at both mu and delta receptors by three orders of magnitude. (c) The caged compound (6) is inactive in comparison to leu-enkephalin (LE).}
\label{fig3}
\end{figure}

Figure~\ref{fig3} illustrates a small sampling of the bounty of newly available tools for the anatomically precise and neuron-type-specific real-time control and measurement of cortical neuropeptide signaling in single-cells in synaptic network context. These and many other emerging effectors and sensors are well-suited to application in fluorescence imaging experiments – conducted in brain slices to optimize analytic interpretation and/or in vivo to optimize physiological and behavioral context. Such experimentation may be guided by coupling matrix predictions like those of figure 2C and 2D or by many other old or new physiological or behavioral hypotheses.

The eight panels of Figure~\ref{fig3} are ordered by stages of NP signaling from peptide release to receptor binding to downstream intracellular signal transduction events. Figure~\ref{fig3}A demonstrates a UV-activatable caged neuropeptide~\cite{banghart2017caged}. Figure~\ref{fig3}B represents a neuropeptide vesicle exocytosis reporter~\cite{ding2019imaging}. Figure~\ref{fig3}C exemplifies similar GPCR-based ligand sensors developed independently by two research groups~\cite{patriarchi2018ultrafast, wu2019g}. Given structural similarities across the parent monoamine- and neuropeptide-selective GPCRs, both groups have validated these sensor platforms for growing numbers of NP-GPCRs. Figure~\ref{fig3}D illustrates results from an imaging application of one of these sensors to visualize dopamine release in vivo~\cite{patriarchi2018ultrafast}. Figure~\ref{fig3}E represents genetically encoded, light-activated GPCRs~\cite{tichy2019light, siuda2015spatiotemporal}. Figure~\ref{fig3}F illustrates a genetically encoded reporter of GPCR activation~\cite{stoeber2018genetically}. Figure~\ref{fig3}G represents design and both cell culture and in vivo imaging applications of a fluorescent-protein-based cAMP sensor~\cite{harada2017red}. Figure~\ref{fig3}H illustrates a protein kinase A activity sensor~\cite{chen2014pka}.

\setcounter{figure}{2}
\begin{figure}[t]
    \renewcommand{\thefigure}{\arabic{figure} (ctd)}
    \caption{\small (d) UV light pulses produce temporally precise suppression of synaptic transmission in slices of hippocampal cortex~\cite{banghart2017caged}.
{\bf B.} (a) Ca-induced-fluorescence principle of a genetically encoded dense-core vesicle exocytosis reporter, Neuropeptide Release Reporter (NPRR). (b) Large increases in fluorescence are observed at larval Drosophila neuromuscular junctions with electrical stimulation of motor nerves~\cite{ding2019imaging}.
{\bf C.} Genetically encoded GPCR-based ligand sensors developed independently by two groups: (a) the “dLight1” dopamine sensor and (b) rapid fluorescence increase electrical stimulation in transfected mouse striatum~\cite{patriarchi2018ultrafast}, and (c) “GRAB” sensors and responses with bath application to transfected HEK293T cells of the ligands indicated~\cite{wu2019g}.
{\bf D.} An imaging application of the genetically encoded dLight1 sensor to visualize dopamine release in vivo~\cite{patriarchi2018ultrafast}.
{\bf E.} (a) Structural chimera design principle of genetically encoded light-activated GPCRs, “opto-XRs”~\cite{tichy2019light}. (b) light stimulation of the “opto-MOR” (rhodopsin/mu-opioid NP-GPCR chimera) produces the reduction of forskolin-induced cAMP production in HEK cells expected for an active mu-opioid receptor~\cite{siuda2015spatiotemporal}.
{\bf F.} (a) Design and (b) use principles of a genetically encoded reporter of native mu-opioid NP-GPCR activation. (c) image format and (d) time-resolved results of opioid ligand and antagonist applications to transfected HEK293 cells~\cite{stoeber2018genetically}.
{\bf G.} (a). Design principle of “pink flamindo”, a genetically encoded, red fluorescent-protein-based single-wavelength cAMP sensor compatible with simultaneous use of a genetically encoded calcium sensor, (b) time-resolved measurements illustrating pharmacologically stimulated anti-phase oscillation of calcium and cAMP, and (c) in vivo imaging of pharmacologically stimulated cAMP increases in pink-flamindo-expressing mouse cortical astrocytes~\cite{harada2017red}. 
{\bf H.} (a) Design principle of a genetically encoded protein kinase A activity sensor, FLIM-AKAR, and (b) results from an experiment using FLIM-AKAR to image cAMP in a single dendritic spine~\cite{chen2014pka}.}
\end{figure}

All of these genetically encoded tools are suitable for use in combination with neuron-type-specific and/or anatomically specific expression and labeling strategies (e.g., via Cre driver lines and/or viral vectors) to explore peptide responses of specific neuron types and to test type-specific coupling predictions like those illustrated in Fig. 2C and 2D. These and many other genetically encoded tools seem certain to lead in the near future to still sharper delineation of neuropeptide involvement in modulating forms of synaptic, membrane and network plasticity like those sampled in Figs. 1C-F. They also seem likely to bring new light to presently enigmatic but critically important issues of neuropeptide secretion and diffusion dynamics in situ within the intricate confines of complex brain tissues.

\section*{Neuropeptides and synaptic learning rules}
Studies of long-lasting forms of plasticity at individual synapses, e.g., long-term potentiation (LTP), long-term depression (LTD) or, more generally, spike-timing-dependent plasticity (STDP), form the foundation for many of today’s models for the formation of memory engrams in synaptic networks~\cite{suvrathan2019beyond, nicola2019diversity, turrigiano2017dialectic}. There is also now a growing body of evidence to suggest that synaptic learning rules (e.g., STDP curves) are themselves subject to parametric adjustment by neuroactive ligands acting via GPCRs~\cite{nadim2014neuromodulation, surmeier2010role}. Such local learning rules, replicated throughout synaptic networks, are now widely hypothesized as “modulated STDP” mechanisms to provide a basis for long-lasting network adaptations and the formation of meaningful memory engrams~\cite{farries2007reinforcement, roelfsema2018control, pang2019fast}. Many lines of evidence, in addition to those already touched upon here, suggest that neuropeptides acting via NP-GPCRs may play central roles in the modulation of synaptic learning rules~\cite{gotzsche2016role, borbely2013neuropeptides}.

Grappling the dynamic peptidergic and synaptic network interactions underlying learning and memory is certain to require theoretical and computational approaches. Work at the presently very active and fertile intersection of synaptic learning studies in neuroscience and machine learning in computer science~\cite{roelfsema2018control, lillicrap2019backpropagation, richards2019deep,srinivasan2018deep, huh2018gradient} is salient here. Efforts in neuroscience and computer science to model or engineer learning in deep neural networks (be they biological or artificial) share the hard problem of individually adjusting very large numbers of “synaptic weights” (so called in both fields) to adapt network function according to experience. This problem is called “credit assignment” because “credit” (or “blame”!) must be assigned correctly to guide strengthening (or weakening) of particular individual synapses dependent upon their contributions to success (or failure) in a given perceptual, mnemonic or motor task. Neuroscientists struggle with credit assignment as they search for rules of synaptic plasticity adequate to explain biological learning. Computer scientists are driven by a quest for greater computational efficiency in training artificial neural networks and the suspicion that evolution may have found better solutions to the credit assignment problem than those conceived so far by human inventors.

The box entitled “Neuronal networks as labeled multidigraphs” delineates a model that schematizes ways that studies of neuromodulatory and synaptic network interplay might use high-resolution neurotaxonomies like those now emerging from single-cell transcriptomes~\cite{fishell2019interneuron, gatto2019neuronal, zeng2017neuronal, tasic2018shared, huang2019diversity, zeisel2018molecular} and now emerging from delineation of synaptic connectomes based on high-throughput electron microscopy~\cite{jonas2015automatic, motta2019dense, seung2014neuronal} and viral tracing methods (e.g.,~\cite{hanchate2020connect}). This multidigraph model, representing a stack of distinct modulatory and synaptic networks pinned together at neuron-type identity nodes, was originally developed in the context of our work on prediction of cortical neuropeptide networks from transcriptomic data~\cite{smith2019single}. Here we represent a slightly more general treatment, potentially applicable to the wide range of broadcast neuromodulators acting via high-affinity GPCRs, including monoamines (dopamine, serotonin, etc.), spillover small-molecule transmitters (glutamate, GABA, ACh) and endocannabinoids in addition to neuropeptides. This neurotaxonomic model might also be abstracted beyond coverage of strictly local volume-diffusion-based regimes to encompass modulatory signaling extended via axons projecting between distant brain regions, while still preserving the notion of addressing via differential gene expression. We expect that such neurotaxonomic models may be critical for both experimental and theoretical attempts to draw lines from modulated STDP mechanisms to credit assignment in biological memory engram formation. We also imagine that models involving analogues to such broadcast modulation of local synaptic learning rules may attract new attention from computer scientists seeking to engineer more efficient credit assignment schemes. 

\hspace*{-1.1cm}\noindent\fbox{%
    \parbox{1.1\textwidth}{%
    \footnotesize
\begin{center}
\vspace{-.2cm}
    { \large Box: Neuronal networks as labeled multidigraphs}
\end{center}
\vspace{-.2cm}
The language and tools of combinatorial graph theory provide a natural framework to model and analyze synaptic and modulatory signalling networks. A {\em labeled multidigraph} is a directed graph where multiple labeled connections between nodes are allowed. These connections can represent different types of directed action between node pairs. The nodes of the graph typically represent the neurons in the network, and the different kinds of actions between the nodes are encoded by the edges between them. We denote by $G=(N, E, W)$  the neuronal network interactions, where $N$ is a set of neurons and $E$ is a set of connections, where each connection is described by $(ij)^k$ representing an edge of type $k$ from node $i$ to node $j$. The type of the edge, $k$, models attributes of a connection such as excitatory, inhibitory, or modulatory, and $W_{ij}^k$ denotes the weight of the connection $(ij)^k$. 

\vspace{.2cm}
{\bf Modeling Dynamics:}
We assume that each neuron has an internal state that is available only to the neuron itself within the network. This internal state may be as simple as the scalar membrane potential in the case of leaky integrate-and-fire neuron models, or in more complex models it may hold a list of cellular properties, including levels or states of modulated target proteins.

\begin{wrapfigure}{r}{0.52\textwidth}
\renewcommand\figurename{\footnotesize Figure}
\vspace{-0.4cm}
  \begin{center}
    \includegraphics[width=0.52\textwidth]{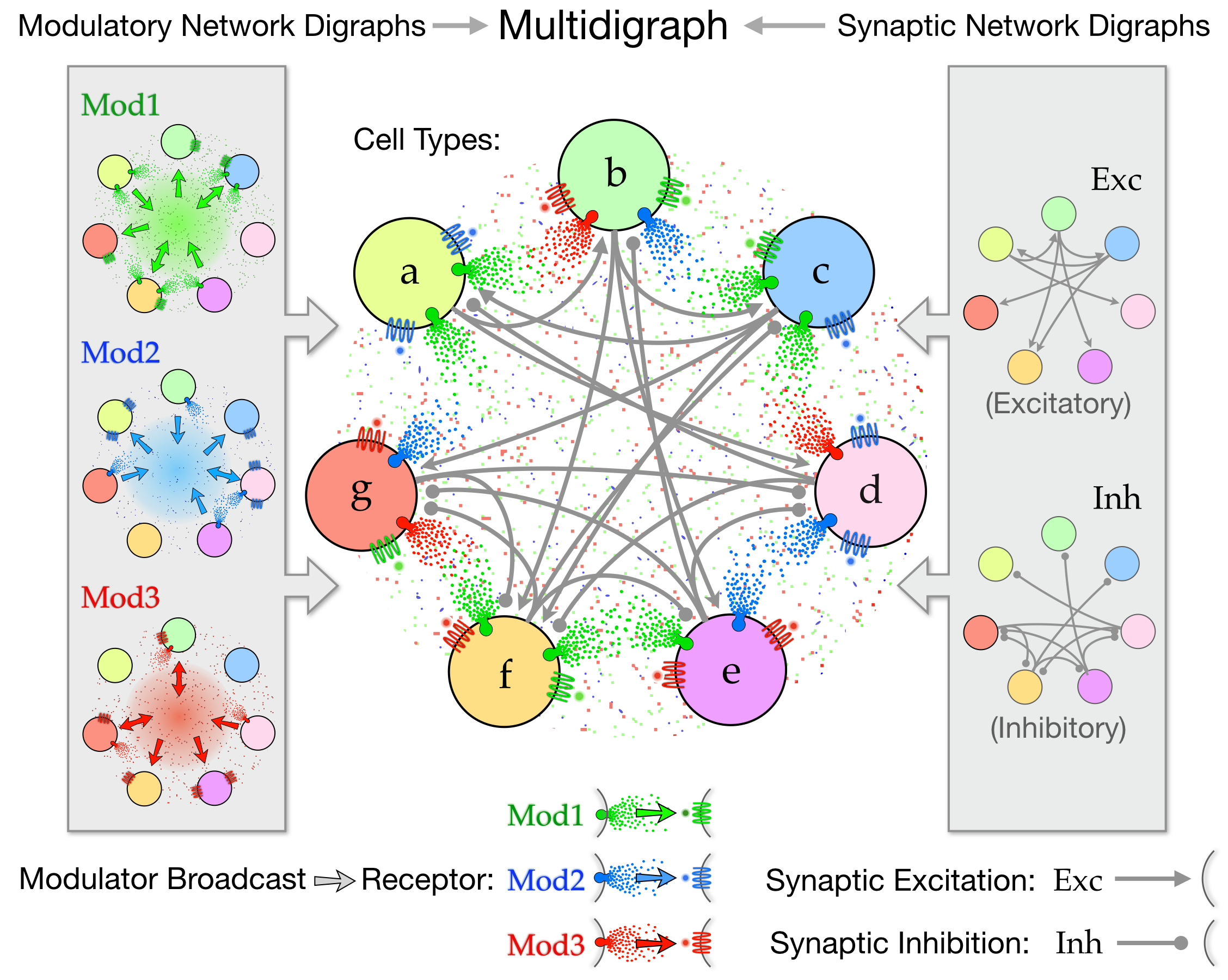}
  \end{center}
  \vspace*{-8mm}
  \caption{\footnotesize {\bf Synaptic and broadcast communication in a neuronal network can be modeled as a multidigraph.} $\alpha,\beta\in\{\mathrm{a},\mathrm{b},\mathrm{c},\mathrm{d},\mathrm{e},\mathrm{f},\mathrm{g}\}$ and $k\in\{\mathrm{Exc},\mathrm{Inh},\mathrm{Mod1},\mathrm{Mod2},\mathrm{Mod3}\}$ denote the cell type and the output type, respectively.}
\end{wrapfigure}

Each neuron generates outputs (e.g., synaptic and modulatory) based on its state, which are observed by other neurons connected to the source neuron by the corresponding edge type. Specifically, output $k$ of neuron $i$ at time $t$, $x_i^k(t)$, is generated by
\begin{equation}
x_i^k(t)=f_k(s_i(t)),
\label{eq_obs}
\end{equation}
where $s_i(t)$ is the state of neuron $i$ at time $t$ and $f_k$ is an output-specific emission function. This output is observed by neuron $j$ through the connection weight $W_{ij}^k$. Let $S=\{\mathrm{Exc}, \mathrm{Inh}\}$ represent synaptic communication so that, when $k\in S$, $W_{ij}^k$ corresponds to the synaptic weight between neurons $i$ and $j$, and $x_i^k$ corresponds to spike output of neuron $i$. Similarly, let $M=\{\mathrm{Mod1}, \mathrm{Mod2}, \mathrm{Mod3},\ldots\}$ be a set of broadcast signaling pathways so that, when $k\in M$, the slower and smoother nature of broadcast communication is reflected in $x_i^k(t)$.

Importantly, we posit here that broadcast modulatory impacts can be modeled neurotaxonomically, that is, as summing across individual cells on a source-cell-type by receiving-cell-type basis, with edge weights set by patterns of source- and receptor-protein gene expression~\cite{smith2019single}. Such neuron-type-level specificity may be ideally suited to theoretical exploration of network homeostasis and plasticity. By expanding upon the approach of Bellec {\it et al}.~\cite{bellec2019solution} and switching to a continuous-time notation, we summarize neuronal dynamics as

\begin{equation}
    \frac{\diff s_j}{\diff t}=\mathcal{F} \Big( s_j,\,
    \underbrace{\vphantom{\sum_{i\in\alpha}} \{(x_i^k, W_{ij}^k): W_{ij}^k \neq 0, k\in S\}}_{\text{synaptic}},\,
    \underbrace{\{( \sum_{i\in \alpha} x_i^k, \mathbf{W}_{\alpha \beta}^k): \mathbf{W}_{\alpha\beta}^k \neq 0, k\in P\}}_{\text{modulatory}},\, u_j \Big),
    \label{eq_dynamics}
\end{equation}
where $s_j$ is the state and we have suppressed the time dependence of the variables and communication delays for simplicity of notation, $\mathcal{F}$ denotes the transformation governing neuronal dynamics, $\alpha$ and $\beta$ denote cell types, neuron $j$ is of type $\beta$, and $u_j$ denotes external input to neuron $j$. In its simplest form, the diffusive nature of broadcast signaling suggests that the weights between cell types can be considered as averages over cells of the same type: $\mathbf{W}_{\alpha\beta}^k=\langle W_{ij}^k \rangle_{i\in\alpha, j\in\beta}$, adding extra structure to a simpler weighted synaptic model.

\vspace{.2cm}
{\bf Temporal credit assignment:} Many real world tasks require action on the time scale of synaptic communication, necessitating efficient learning of $W_{ij}^\mathrm{Exc}$ and $W_{ij}^\mathrm{Inh}$, and suggesting an auxiliary role for modulatory signaling (and $W_{ij}^k$, $k\in P$). Inspired by results implicating dopamine modulation of spike-timing-dependent plasticity (STDP) in credit assignment~\cite{roelfsema2018control,fremaux2016neuromodulated, miconi2020backpropamine}, we hypothesize that other forms and patterns of broadcast modulatory signaling may also promote efficient training of spiking networks. While an overarching temporal credit assignment theory for multidigraphs does not yet exist, approaches enabling credit assignment at multiple time scales~\cite{bellec2019solution,miconi2020backpropamine,neftci2019surrogate,bellec2018long} can provide natural solutions for incorporating the rich and flexible repertoire of broadcast modulation. Finally, devising approximate, surrogate gradients~\cite{huh2018gradient,bellec2019solution,neftci2019surrogate,zenke2018superspike} for the non-differentiable spiking signal enables efficient in-silico experiments and brings the fields of computational neuroscience and machine learning closer.
}
}
\section*{Conclusions}
Single-cell RNA-seq transcriptomes are now offering new and surprising perspectives on cortical neuropeptide signaling. While the likelihood of intracortical neuropeptide signaling was recognized some time ago (e.g.,~\cite{gallopin2006cortical}) and it has long been known that dozens of NP signaling genes are expressed in cortex, it now appears that all cortical neurons express at least one NPP and at least one NP-GPCR gene. Transcriptomic and anatomical data now strongly suggest that local broadcast of NP products interconnects every cortical neuron by a tall stack of overlapping neuron-type-specific peptidergic modulator networks. Patterns of NP gene expression are moreover highly coherent with the deep neurotaxonomies now materializing from genome-wide single-cell transcriptomes. This coherence joins with new opportunities for experimental access offered by transcriptomic neurotaxonomy to make transcriptomic predictions of cortical NP networks highly amenable to experimental test and theoretical exploration. The coherence of transcriptomic neuron type with NP signaling gene expression also suggests that connections between the differentiation of neuron types and differentiation of neuropeptide networks are ancient and fundamental to brain function.

Many longstanding open questions about roles of neuropeptide signaling in the functional homeostasis and plasticity of cortical synaptic networks are re-framed by these new transcriptomic perspectives. In this writing, we have attempted to highlight new results that frame old questions into new hypotheses and to touch upon fundamental advances and new tools that are now making such hypotheses specific and testable. We look forward very eagerly to watching this progress promote better understanding of the homeostasis, adaptation and memory-related plasticity of cortical synaptic networks, and to seeing what new principles of cortical computation such understanding brings from present dark corners into the light!

\section*{Acknowledgements}
The authors are grateful to Drs. Scott Owen, Rohan Gala, Forrest Collman and Christof Koch, and to Leila Elabbady for many helpful discussions and comments.

\section*{Funding}
We wish to thank the Allen Institute founder, Paul G. Allen, for his vision, encouragement and support.

\bibliography{newlight}
\bibliographystyle{unsrt}
\end{document}